\documentclass[12pt]{amsart}
\usepackage{amsmath,amsfonts,amssymb} \usepackage{color}
\usepackage[all,cmtip]{xy}
\usepackage{graphicx}
 \usepackage{youngtab}

\newcommand{\be}{\begin{equation}}
\newcommand{\ee}{\end{equation}}
\newcommand{\bea}{\begin{eqnarray}}
\newcommand{\eea}{\end{eqnarray}}
\newcommand{\ben}{\begin{eqnarray*}}
\newcommand{\een}{\end{eqnarray*}}

\newtheorem{cor}{Corollary}

\newtheorem{conj}[cor]{Conjecture}

\theoremstyle{remark}

\definecolor{A}{rgb}{.75,1,.75}

\definecolor{green}{rgb}{0,1,0}
\definecolor{yellow}{rgb}{1,1,0}
\definecolor{orange}{rgb}{1,.7,0}
\definecolor{red}{rgb}{1,0,0}
\definecolor{white}{rgb}{1,1,1}

 \makeindex
\begin{document}
\title
{Frobenius Manifolds, Spectral Curves, and Integrable Hierarchies}

\author{Jian Zhou}
\address{Department of Mathematical Sciences\\Tsinghua University\\Beijng, 100084, China}
\email{jzhou@math.tsinghua.edu.cn}

\begin{abstract}
We  formulate some conjectures that relates semisimple Frobenius manifolds,
their spectral curves and integrable hierarchies.
\end{abstract}

\maketitle

In this short note we will propose some conjectures to unify the reconstruction approach and the emergent
approach in the study of Gromov-Witten type theories.

As explained in an influential article by Anderson \cite{Anderson}
and an inspiring book by Laughlin \cite{Laughlin},
there are the reduction/reconstruction and the emergence approaches to science.
In the former approach one tries to derive the collective behavior of large quantity of individual particles
from the fundamental laws obeyed by each individual particle.
In this approach the difficulty is that one has to first find the fundamental laws,
and secondly the complexity of deriving the collective behavior from these laws makes it very difficult
and often out of reach.
In the latter approach there are fundamental laws at each level of complexity \cite{Anderson},
and it is even possible that all the fundamental laws for individual particles
have their origins in their collective behavior \cite[Preface, XV]{Laughlin}.
So fundamental laws might emerge from the collective behavior.

In some recent papers \cite{Zhou-Quant-Def, Zhou-Emergent, Zhou-KP, Zhou-KP-II},
we have embarked on a study of Gromov-Witten type theories by an emergent approach.
A Gromov-Witten type theory is roughly speaking a topological field theory that describes topological matters
coupled to 2D topological gravity.
The topological matters give us a finite-dimensional space (called the small phase space)
of topological observables (called the primary observables).
When coupled to the topological gravity,
each observable has an infinite sequence of gravitational descendants.
One then gets an infinite-dimensional space called the big phase space.
A Gromov-Witten type theory defines a partition function on the big phase space.
A reconstruction approach towards a Gromov-Witten type theory is to reconstruct
the partition function on the big phase from the restriction of its genus zero part
on the small phase space.

In many examples,
the genus zero free energy partition of a Gromov-Witten type theory
satisfies the WDVV equations.
A geometric structure called a Frobenius  manifold was introduced by Dubrovin \cite{Dubrovin}
to encode such information.
It forms the foundation of some reconstruction approaches and plays the role of fundamental laws of
individual particles.
Two formalisms have been proposed to reconstruct the whole theory
from a Frobenius manifold,
under a technical condition of semisimpleness.
One of them is Givental's quantization formalism \cite{Givental}.
It takes a product of finite many copies of the partition function of the Gromov-Witten theory of a point,
the number of which equals the number of primaries,
and obtains from it the partition function of the whole theory by the action of some operators
obtained from quantization of a symplectic structure on the formal loop space of
the small phase space.
The other formalism due to Dubrovin-Zhang \cite{Dubrovin-Zhang} is to first extend the system of WDVV equation
to  an infinite system of equations called the principal hierarchy.
This hierarchy is regarded as the dispersionless limit (i.e. the genus zero part)
of an integrable hierarchy satisfying certain axioms,
called the Dubrovin-Zhang hierarchy.
This is inspired by Krichever's observation \cite{Kri-Dispersionless}
that certain solutions of dispersionless limit of integrable hierarchies lead to solutions of WDVV equations.
Since the Dubrovin-Zhang hierarchy is some special deformation of the principal hierarchy,
one can try to establish its existence
and uniqueness by computing certain cohomology groups \cite{Liu-Zhang,CPS}.
One can also establish the equivalence of these two formalisms \cite{Liu}.
Both the Givental formalism and the Dubrovin-Zhang formalism are inspired by Witten Conjecture/Kontsevich Theorem
\cite{Wit, Kon}.
Whereas in Dubrovin-Zhang formalism integrable hierarchy plays a key role,
connections to integrable hierarchies can also be made in the Givental formalism
in many examples \cite{Givental, Giv-Mil, Fre-Giv-Mil, Fan-Jarvis-Ruan, Liu-Ruan-Zhang}.

A different reconstruction approach, called Eynard-Orantin topological recursion \cite{Eyn-Ora},
starts with a plane curve with some extra data
and defines $n$-point functions recursively.
In appearance this looks quite differently from either the Givental formalism or the Dubrovin-Zhang formalism,
since both of them starts with a semisimple Frobenius manifold.
Nevertheless
it has been shown that a local version of topological recursion is equivalent to Givental's quantizantion formalism \cite{DOSS}.
This formalism turns out to be very successful in reformulating local mirror symmetry
of toric Calabi-Yau 3-folds \cite{BKMP}.
Both the proofs in the case of toric Calabi-Yau 3-manifold \cite{Eyn-Ora-BKMP}
and 3-orbifolds \cite{Fang-Liu-Zong} use comparison with recursions
in the Givental formalism.
By these results,
one may expect to establish some connections between the global 
EO topological recursions and the Frobenius manifold theory by
associating spectral curves to Froebnius manifolds.
In fact,
Dubrovin \cite{Dubrovin} defined a superpotential function of a semisimple Frobenius manifold
very long ago\footnote{The author thanks Professor Si-Qi Liu for bringing this to his attention a couple of years ago.},
and people suspect that the partition function of the semisimple Frobenius manifold
obtained by Givental formalism or Dubrovin-Zhang formalism satisfies
the Eynard-Orantin topological recursion using the spectral curve defined by
this superpotential function.
This has been shown to be the case \cite{DNOPS}.

One can consider the problem of going now in the reversed direction,
i.e., 
from topological recursions on a spectral curve to Frobenius manifold.
A consequence of being able to do so is that one can clearly see that 
a tau-function of some integrable hierarchy can be produced in this way. 
As pointed out by Eynard and Orantin \cite{Eyn-Ora},
the partition function defined by the topological recursion should be the tau-function
of some integrable hierarchy.
This was further elaborated by Borot and Eynard \cite{Borot-Eynard}.
 
We will refer to Givental formalism and Dubrovin-Zhang formalism as Type A reconstruction formalisms,
to Eynard-Orantin formalism as a Type B reconstruction formalism.
As we point out above,
both types of reconstruction formalisms lead to integrable hierarchies,
this is why we take integrable hierarchies as our foundation for an emergent approach.
We take the point of view of the Kyoto school,
i.e.,
we treat general integrable hierarchies as reductions of KP or 
n-component KP hierarchies \cite{Sato, Jimbo-Miwa}.
The KP case has been studied in \cite{Zhou-KP, Zhou-KP-II},
the n-component KP \cite{Kac-van} case will be reported in \cite{Zhou-n-KP},
and some reductions will be reported in \cite{Zhou-Reduction}.
The main results of these work can be summarized as follows.
Given a tau-function of one of these integrable hierarchies,
one can produce a spectral curve by studying the Kac-Schwarz operators
that specify it,
together with some natural quantization and deformation over the big phase space.
One can derive from this the W-constraints satisfied by the tau-function.
The quantization of the spectral curve leads to a reduction of the integrable
hierarchy to
a dispersive universal Whitham hierarchy introduced by
Tanisaki-Takebe \cite{Tak-Tak} and
Szablikowski-B{\l}aszak \cite{Sza-Bla} associated with the spectral curve in genus zero.
Recall the dispersionless universal Whitham hierarchy was introduced by Krichever \cite{Kri-Whitham}
for a curve of arbitrary genera.

This short note is driven by the hope that  by putting all these results together,
one can find a unification of different approaches.
In the case at hand,
the reconstruction approaches and the emergent approach
are just two sides of the same coin.
They nicely fit together as they should.
This can be best seen if
we summarize the work mentioned above about
the relationship among three subjects in the title
in the following picture:
$$
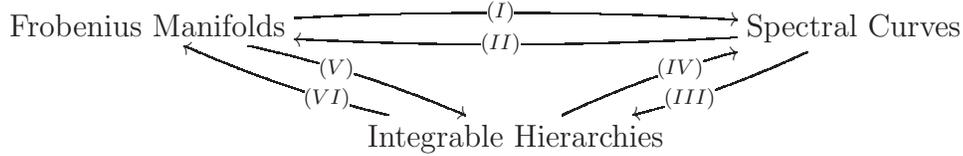
\xymatrix{
\text{Frobenius Manifolds} \ar@/^/[rr]|{(I)} \ar@/^/[dr]|{(V)}
                &  &    \text{Spectral Curves} \ar@/^/[dl]|{(III)} \ar@/^/[ll]|{(II)}    \\
                & \text{Integrable Hierarchies}  \ar@/^/[ru]|{(IV)}  \ar@/^/[lu]|{(VI)}             }
$$
Here we content ourselves with semisimple Frobenius manifolds.
Links in the top row can be thought of as mirror symmetry,
links from the top row to the bottom row are the reconstruction approaches,
and the links from the bottom row to the top row can be regarded as
emergent approaches.
The link (I) is established in \cite{DNOPS, DOSS}.
A direct link (II) is missing,
an indirect link is through (III) and (VI).
For (III), see \cite{Eyn-Ora, Borot-Eynard} and this paper.
We establish (IV) for KP and n-component KP hierarchies in \cite{Zhou-KP}
and \cite{Zhou-n-KP} respectively.
For examples of (VI), see \cite{Kri-Whitham, Dub-Whitham}.
For examples of (V), see \cite{Dubrovin-Zhang, Giv-An, Giv-Mil, Fre-Giv-Mil,
Fan-Jarvis-Ruan, Liu-Ruan-Zhang}.

To strengthen the above unification of reconstruction and emergent approaches,
let us now formulate some conjectures.

\begin{conj}
One can generalize Krichever's construction of
universal Whitham hierarchy on punctured Riemann surfaces of arbitrary genera
to the dispersive case.
\end{conj}

\begin{conj}
By Eynard-Orantin topological recursion one can construct
a tau-function of the dispersive universal Whitham hierarchy
associated with the spectral curve.
\end{conj}

By combining the above two conjectures with the construction
of spectral curve of semisimple Frobenius manifold \cite{DNOPS},
we make the following:

\begin{conj}
The Dubrovin-Zhang integrable hierarchy associated with a semisimple Frobenius manifold
can be identified with the universal Whitham hierarchy associated
to the spectral curve of the Frobenius manifold constructed
using Dubrovin's potential function.
\end{conj}

There are many different theories that produces semisimple Frobenius manifolds,
so our conjecture proposes the integrable hierarchy satisfied by their partition functions.
For example,
Gromov-Witten theory of Fano manifolds,
Saito's theory of primitive forms, etc.
A particularly interesting application is to deformation theory of isolated singularities.
One first uses Saito's theory to produce a semisimple Frobenius manifold,
then uses Dubrovin's superpotential to produce a spectral curve,
our conjecture then produces a tau-function of the integrable hierarchy 
associated to the spectral curve.
Emergent geometry of the integrable hierarchy then produces a quantum deformation 
theory of the integrable hierarchy.

By combining Conjecture 1 with the proof of the BKMP Remodelling Conjecture
\cite{Eyn-Ora-BKMP, Fang-Liu-Zong},
we make the following:

\begin{conj}
Partition function of a toric Calabi-Yau 3-manifold or 3-orbifold with
Aganagic-Vafa outer D-branes is a tau-function of
universal Whitham hierarchy associated with its local mirror curve.
\end{conj}

One can also compare Conjecture 3 and Conjecture 4 and make a connection
between them.
It was conjectured in \cite{ADKMV} that the partition functions in Conjecture 4
are tau-functions of n-component KP hierarchy and this leads to
a fermionic reformulation of the theory.
Some special cases have been verified in \cite{Zhou-Int}.
For more partial results,
see also the more recent work \cite{Deng-Zhou1, Deng-Zhou2}.

\begin{conj}
There is a fermionic reformulation of dispersive universal
Witham integrable hierarchy associated with any spectral curve.
\end{conj}

\begin{conj}
There is a fermionic reformulation of the Dubrovin-Zhang
integrable hierarchy for semisimple Frobenius manifolds.
\end{conj}

Of course there is no reason that stops us from conjecturing that
any Gromov-Witten type theory has a fermionic reformulation.

\vspace{.2in}

{\em Acknoledgements}.
This research is partially supported by NSFC grant 11171174.
The author thanks Professors Si-Qi Liu and Youjin Zhang
for explaining to him the Dubrovin-Zhang theory
and its relationship to Givental formalism.
The author also thanks them and Professor Yongbin Ruan
for explaining their joint work.

\end{document}